
\long\def\UN#1{$\underline{{\vphantom{\hbox{#1}}}\smash{\hbox{#1}}}$}
\magnification=\magstep 1
\overfullrule=0pt
\hfuzz=16pt
\voffset=0.0 true in
\vsize=8.8 true in
\baselineskip 20pt
\parskip 6pt
\hoffset=0.1 true in
\hsize=6.3 true in
\nopagenumbers
\pageno=1
\footline={\hfil -- {\folio} -- \hfil}

\centerline{{\bf Anisotropic
Diffusion-Limited}}

\centerline{{\bf Reactions with Coagulation and
Annihilation}}

\vskip 0.4in

\centerline{{\bf Vladimir Privman},\ \ {\bf Ant\'{o}nio
M.~R.~Cadilhe},\ \ and\ \ {\bf M.~Lawrence Glasser}}

\vskip 0.2in

\centerline{\sl Department of Physics, Clarkson University,
Potsdam, New York 13699--5820, USA}

\vskip 0.4in

\centerline{\bf ABSTRACT}

One-dimensional reaction-diffusion models $A+A \to
{\emptyset}$, $A+A \to A$, and $A+B \to {\emptyset}$, where
in the latter case like particles coagulate on
encounters and move as clusters, are solved exactly with
anisotropic hopping rates and assuming synchronous
dynamics.  Asymptotic large-time results for particle
densities are derived and discussed in the framework of
universality.

\


\vfil\eject

Diffusion-limited reactions involving aggregation and
annihilation processes are important in many physical,
chemical and biological phenomena [1] such as star
formation, polymerization, recombination of charge carriers
in semiconductors, soliton and antisoliton annihilation,
biologically competing species, etc. In this work we study
by exact solution effects of anisotropy in $1D$, for
single-species reactions $A+A\to A$ or ${\emptyset}$, and a
two-species annihilation model $A+B\to{\emptyset}$ in which
like particles coagulate irreversibly.  Scaling approaches
[1-2] suggest that in $1D$ these reactions are
fluctuation-dominated, and we cannot expect the rate
equation approach to be valid. Indeed the mean-field rate
equation approximation ignores effects of inhomogeneous
fluctuations. Exact solutions and asymptotic arguments, in
$1D$, have been used [3] to check general scaling and
universality expectations. The $1D$ reactions have also
found some experimental applications [4]. These studies
have assumed isotropic hopping (diffusion).

For the reaction $A+B\to{\emptyset}$, numerical results and
phenomenological considerations suggest [5] that making the
hopping fully directed would change the universality class
in $1D$. Specifically, the large-time particle
concentrations (assuming equal densities of both species)
would scale according to $c (t) \sim t^{-1/3}$ instead of
the isotropic-hopping power law $t^{-1/4}$. A few exact and
numerical results available in the literature on
anisotropic reactions involving only one species [6]
indicate that the power law is not changed. The model of
[5] assumed that like particles interact via hard-core
repulsion; this seems to be an essential ingredient for
observing the changeover in the universality class.

In this work we report the exact solution for two-particle
annihilation with anisotropic hopping.  However, in order
to achieve exact solvability we took ``sticky-particle''
rather than hard-core interactions: the like particles
coagulate on encounters and diffuse as groups. Our exact
calculations yield the $t^{-1/4}$ power law, found earlier
for ``sticky-particles'' with different dynamics and
isotropic hopping [7].  For unequal initial concentrations,
the large-time behavior changes; the crossover between the
two regimes is derived analytically. We also obtain exact
results for $A+A\to A$ or ${\emptyset}$ with anisotropy.
The universality class of dynamics of these reactions is
not affected by hopping anisotropy. An extended version of
this work will be reported elsewhere [8].

In lattice models the particles hop randomly, to the extent
allowed by their interactions, to their nearest neighbor
sites. Two like particles can annihilate on encounters,
$A+A\to {\emptyset}$, or aggregate, $A+A\to A$.  The $1D$
kinetics of these reactions is non-mean-field, with the
typical large time diffusional behavior of the
concentration (density per site), $c (t) \sim t^{-1/2}$.
For the two-species model, to be termed the $AB$ model,
unlike particles annihilate, $A+B\to {\emptyset}$. When
like species meet, some interaction must be assumed. The
simplest interaction is hard-core.  Assuming equal $A$- and
$B$-concentrations and random, uniform initial conditions,
particle concentrations in the isotropic case scale
according to $c (t) \sim t^{-1/4}$ in $1D$. A surprising
recent result [5] is the new exponent $\approx 1/3$,
replacing $1/4$, for anisotropic \UN{\it hard-core}\
particle hopping.

In order to obtain a solvable model in $1D$, we consider
the $AB$ annihilation model with the ``sticky particle''
interaction. Thus, like particles coagulate irreversibly on
encounters, e.g., $nA+mA\to (n+m)A$, and diffuse as
clusters.  When unlike clusters meet at a lattice site, the
outcome of the reaction is $ nA+mB \to (n-m)A $ if $n>m$,
$\emptyset $ if $ n=m $, and  $(m-n)B $ if $ n<m$.  Recent
numerical results and scaling considerations for these
reactions [9] in $D=1,2,3$ indicate that they are
mean-field in $D=2,3$. However, in $1D$ the power-law
exponent for the density is $1/4$ [7,9], with a faster
power-law decay $\sim t^{-3/2}$ of the minority species in
case of unequal densities of $A$ and $B$.

Following [10], we first consider diffusion of \UN{\it
nonnegative}\  charges on the $1D$ lattice. Initially, at
$t=0$, we place positive unit charge at each site with
probability $p$ or zero charge with probability $1-p$.
Furthermore, we consider synchronous dynamics, i.e.,
charges at all lattice sites hop simultaneously in each
time step $t \to t+1$, where the probabilities of hopping
to the right, $r$, and to the left, $\ell=1-r$, are not
necessarily equal. This dynamics decouples the even-odd and
odd-even space-time sublattices; it suffices to consider
only those charges which are at the lattice sites $j=0,\pm
2,\pm 4, \ldots$ at times $t=0,2,4,\ldots$, and lattice
sites $j=\pm 1, \pm 3 , \pm 5, \ldots$ at times $t=1,3,5,
\ldots$.  The ``interaction'' between the charges is
defined by the rule that all charge accumulated at site $j$
at time $t$ coagulates. There can be 0, 1 or 2 such charges
arriving at $j$, depending on the random decisions
regarding the directions of hopping from sites $j \pm 1$.

This model can also be viewed as diffusion-coagulation of
unit-charge ``particles'' $C$, i.e., $nC+mC\to (n+m)C$.
Such reactions, without the limitation of positive or
integer charges, and with an added process of feeding-in
charge at each time step, have been considered as models of
self-organized criticality and coagulation [11-12],
assuming isotropic hopping, $r=\ell={1 \over 2}$. This
coagulation reaction can be mapped [7,10] onto both
our single-species and ``sticky'' $AB$ models.  However,
before discussing this mapping, let us present the exact
solution of the model of coagulating charges with
anisotropic hopping, following the ideas of [10-11].

We define stochastic variables, $\tau_j (t)=1$ or $0$, with
probabilities $r$ and $\ell$, respectively.  The stochastic
equation of motion for the charges $q_j (t)$, equal to the
number of $C$ particles at site $j$ at time $t$, is $$q_n
(t+1) = \tau_{n-1} (t) q_{n-1} (t) + \left[ 1-\tau_{n+1}
(t) \right] q_{n+1} (t) \;\; . \eqno(1)$$ The total number
of $C$-particles, or the total charge, in an interval of
$k$ consecutive proper-parity-sublattice sites, starting at
site $j$ at time $t$, is given by $$S_{k, j}
(t)=\sum_{i=0}^{k-1} q_{j+2i} (t)=q_j (t)+q_{j+2}
(t)+\cdots+ +q_{i+2k-2} (t) \;\; . \eqno(2)$$ Due to
conservation of charge, the equations of motion (1) yield
the following relation, $$ S_{k, n} (t+1)=\tau_{n-1} (t)
q_{n-1} (t) +q_{n+1} (t)+\cdots+q_{n+2k-3} (t) +
\left[1-\tau_{n+2k-1} (t)\right]q_{n+2k-1} (t) \;. \;(3) $$
Thus, only the two random decisions at the end points are
involved in the dynamics of charges in consecutive-site
intervals. The exact solvability of coagulating-charge
models is based on this property [11].

Let us introduce the function $ I(s,m) = \delta_{s , m} $,
and averages, $ f_{k, m} (t) = \langle I \left( S_{k,
j}(t),m \right) \rangle$.  The averaging is over the
stochastic dynamics, i.e., over $\tau_i (t)$, as well as
over the initial conditions. Since the latter are uniform,
$f_{k, m} (t)$ do not depend on $j$. Other choices for
$I(s,m)$ have been used [10-12].  In our case $f_{k, m}
(t)$ correspond to the probability to find $m$ charge units
in an interval of $k$ sites, so that $f_{1, m} (t)$ is the
density (fraction) of sites with charge $m$.

The variables $\tau_i (t)$ and $S_{k, n} (t)$ are
statistically independent because the latter depend
only on ``decision making'' at earlier times. As a result,
one obtains, by (3), the discrete diffusion-like equation,
$$ f_{k,m} (t+1) =r\ell\left[f_{k+1,m} (t)+f_{k-1,m}
(t)\right]+\left(r^2+\ell^2\right)f_{k,m} (t) \;\; .
\eqno(4)$$ The $m$-dependence only enters via the initial
conditions:
$f_{k,m}(0)=p^m (1-p)^{k-m} {k\choose m}$, provided $0\leq
m\leq k$, and 0 for $m>k$.
We also define the boundary conditions
$f_{0,m} (t) = I(0,m)=\delta_{0,m}$ in order to extend (4)
to all $t =0,1,2,\ldots$.

In order to solve (4) we introduce the double generating
function, $ g_k (u,w ) = \sum_{t=0}^\infty
\sum_{m=0}^\infty f_{k,m}(t) u^t w^m $. It is also
convenient to introduce the variable $a=r-\ell$ directly
measuring the hopping anisotropy, $ r=(1+a)/2$ and $
\ell=(1-a)/2 $.  One can then derive the following
equations, $$g_{k+1}(u,w)+2{\left(1+a^2\right)u-2 \over
\left(1-a^2\right)u}g_k(u,w) +g_{k-1}(u,w)=-{4 \over
\left(1-a^2\right)u}(w p+1-p)^k \;\; , \eqno(5)$$ with the
initial and boundary conditions $g_k(0,w)=(w p +1-p)^k$ and
$g_0(u,w)={1 / ( 1-u)} $.

The solution of (5) is obtained as a linear combination of
the special solution $ \Omega (w p+1-p)^k $ proportional to
the right-hand side, and that solution of the homogeneous
equation which is regular at $u=0$. The coefficient $\Omega
$ is obtained by substitution, $$\Omega =-{4(w p+1-p) \over
\left(1-a^2\right)u \left(w p+1-p-\Lambda_+\right)\left(w
p+1-p-\Lambda_-\right)} \;\; , \eqno(6)$$ where $\Lambda_\pm$
are the roots of the characteristic equation,
$$\Lambda_\pm={2-\left(1+a^2\right)u\pm
2\sqrt{(1-u)\left(1-a^2u\right)} \over \left(1-a^2\right)u}
\;\; . \eqno(7)$$ The root $\Lambda_-$, which is
nonsingular as $u \to 0$, gives the homogeneous solution
proportional to $ \Lambda_-^k$, where the proportionality
constant is determined by the boundary conditions. In
summary, the solution takes the form $$g_k(u,w)=\left({1
\over 1-u}-\Omega \right)\Lambda_-^k+\Omega (w p+1-p)^k
\;\; . \eqno(8) $$

Densities of reactants at lattice sites derive from
$f_{k=1,m}(t)$. The $m$-dependence here follows by
expanding (8) in powers of $w$. The resulting
$u$-dependence is complicated.  Therefore we will keep the
time-dependence in the generating-function form. Our
explicit time-dependent expressions will be derived as
asymptotic results valid for large times. The power series
in $u$ are then controlled by the singularity at $u=1$, and
analytical results can be derived by appropriate
expansions. We consider the time-generating function for
the quantities $f_{1,m}(t)$ which represent the probability
to find charge $m$ at a lattice site at time $t$,\ $$ G_m
(u) = \sum_{t=0}^\infty f_{1,m}(t) u^t
=\delta_{m,0}\left[{\Lambda_- \over 1-u}-{4 \over
\left(1-a^2\right)u} \right]- {4\Lambda_+(-p)^m \over
\left(1-a^2\right)u\left(1-p-\Lambda_+\right)^{m+1}} \;
. \; (9)$$

We now turn to the single-species reactions introduced
earlier. Our approach follows recent work [10] and related
ideas, e.g., [13]. Consider first the reaction $A+A\to A$.
In the coagulating-charge model we now regard each
``charged'' site as occupied by an $A$-particle, and each
``uncharged'' site as empty of $A$-particles. The dynamics
of the coagulating charges then maps onto the dynamics of
the reaction $A+A\to A$. The quantity $f_{1,0}(t)$ gives
the density of empty sites in both models. Therefore, the
particle density (per lattice site), $c(t)$, in the
aggregation model, is given by $ c(t)=1-f_{1,0}(t)$, where
$c(0)=p$.  The generating function follows from (9), $$
E(u)= \sum_{t=0}^\infty c(t)u^t={1 \over 1-u}-G_0(u) ={1
-\Lambda_- \over 1-u}+ {4(1-p) \over
\left(1-a^2\right)u\left(1-p-\Lambda_+\right)} \;\; .
\eqno(10)$$

The function $E(u)$ is regular at $u=0$; the Taylor series
is controlled by the singularity at $u=1$, $$ E(u) = {2
\over \sqrt{1-a^2}} \left[{1 \over \sqrt{1-u}}+{\cal O}(1)
\right] \;\; . \eqno(11)$$ This yields the leading-order
large-time behavior, $$ c(t) \approx {2 \over
\sqrt{\left(1-a^2\right)\pi t\,}} \;\; . \eqno(12) $$ We
are not aware of other exact solutions for this model with
anisotropic hopping. However, the leading-order large time
behavior is expected to be universal in that it does not
depend on the initial density $p$. Furthermore, the
particle diffusion constant ${\cal D}(a)= \left( 1- a^2
\right) {\cal D} (0)$ decreases proportional to $1-a^2$
when the anisotropy is introduced. Therefore, as a function
of ${\cal D}(a)t$, the result (12) does not depend on the
anisotropy and in fact it is the same as expressions found
for other $A+A\to A$ models, with different detailed
dynamical rules [3].

For the reaction $A+A\to {\emptyset}$, the appropriate
mapping is to identify odd charges with particles $A$ and
even charges with empty sites [10]. The generating function
is obtained as follows, $$ E(u) = \sum_{j=0}^\infty
G_{2j+1}(u)= {4\Lambda_+ p \over \left(1-a^2\right)u
\left[\left(1-p-\Lambda_+\right)^2-p^2\right]} \;\; .
\eqno(13)$$ The large-time behavior is similar to the
aggregation reaction, with the universal expression which
is less than (12) by a factor of\ 2.

The finite-time results for both models do depend on
details of the dynamical rules. For our particular choice
of synchronous dynamics, there exists an exact relation [10]
which holds also for the anisotropic case, checked by
comparing the generating functions, $$ 2c_{\emptyset}
(t;p)=c_A (t;2p) \;\; . \eqno(14) $$ Here the subscripts
denote the outcome of the reaction while the added argument
stands for the initial density.

For the $AB$ model, we assume that initially particles are
placed with density $p$, but now a fraction $\alpha$ of
them are type $A$, and a fraction $\beta=1-\alpha$ are type
$B$.  The concentration difference is constant during the
reaction; it remains $(\alpha-\beta) p$. At large times,
this is also the limiting value of the density of the
majority species, while the density of the minority species
vanishes. In what follows we assume $\alpha \geq \beta $
without loss of generality; $c(t)$ will refer to the
density of the majority species $A$.

The dynamics of the $AB$ model can be related to that of
the coagulating-charge model by adapting the ideas of [7].
The dynamics of the ``sticky'' $A+B \to {\emptyset}$ model
can be viewed as coagulation. Thus, we consider the $AB$
particles as new charges, $+1$ for $A$, and $-1$ for $B$.
If the net charge of a coagulated cluster is positive than
we regard it as a group of $A$ particles (equal in their
number to the charge value). If the charge is negative, we
consider the cluster $B$-particle, while if the charge is
$0$, we regard this cluster as nonexistent ($\emptyset$).

The probability of having an $m$-particle cluster in the
original positive-charge-only model was given by
$f_{1,m}(t)$. Each such cluster can have charge $n = -m,
-m+2, \dots, m-2, m$, where we now refer to the new, $\pm$
charge definition. The key observation is that having a
``species'' label assigned to a particle at time $t=0$ is
statistically independent of its motion and coagulation as
part of clusters at later times.  The density (per site) of
$m$-particle clusters with exactly $n$ units of charge can
be calculated as follows,
$$\Psi_{m,n}(t)={\alpha\vphantom{T}}^{m+n \over
2}{\beta\vphantom{T}}^{m-n \over 2} {m! \over \left({ m+n
\over 2}\right)!\, \left({ m-n \over 2}\right)!} f_{1,m}(t)
\;\; . \eqno(15)$$ The concentration of $A$-particles,
i.e., the density per site of the $+$ charge, can be
written as $ c(t)=\sum_{n=1}^\infty n \left[
\sum_{m=n,n+2,\ldots}\Psi_{m,n}(t)\right]$.  After some
algebra, we get the generating function, $$
E(u)={4\Lambda_+ \over \left(1-a^2\right)u\left(p+\Lambda_+
-1 \right)} \left(x{\partial \over \partial x} -y{\partial
\over \partial y}\right) S(x,y) \;\; . \eqno(16) $$ Here we
introduced the function $$ S(x,y)= \sum_{n=1}^\infty \,
\sum_{j=0}^\infty \, x^{n+j} y^{j}{n+2j \choose j} = {2x
\over \sqrt{1-4xy\,} \left(1-2x+\sqrt{1-4xy}\right)} \;\; ,
\eqno(17)$$ and the variables $ x=p\alpha / \left(
p+\Lambda_+ -1 \right) $, $ y=p\beta / \left( p+\Lambda_+
-1 \right) $.  The evaluation of the double-sum is quite
nontrivial; see [8] for details.

It is useful to introduce the parameter $b=\alpha-\beta
\geq 0$ which measures the excess of $A$ at time $t=0$, $$
\alpha=(1+b)/2 \;\;\;\;\;\; {\rm and} \;\;\;\;\;\;
\beta=(1-b)/2 \;\; . \eqno(18)$$ For the equal
concentration case, $b=0$, the large-time behavior is
governed by the singularity at $u=1$, $$E(u) = {1 \over
(1-u)^{3/4}}\left[{\sqrt{p} \over
2\left(1-a^2\right)^{1/4}} -{1-p \over
4\sqrt{p}\left(1-a^2\right)^{3/4}}(1-u)^{1/2}+{\cal
O}(1-u)\right] \;\; . \eqno(19)$$ The leading-order
behavior of the density follows from the first term, $$c
(t) \approx {\sqrt{p} \over
2\Gamma(3/4)\left(1-a^2\right)^{1/4} t^{1/4}} \;\; .
\eqno(20)$$ The most significant feature of this result is
that, similar to the single-species reactions, the
anisotropy, $a$, dependence can be fully absorbed in the
diffusion constant, in terms of ${\cal D}(a)t =
\left(1-a^2\right) {\cal D}(0)t$. The exponent $1/4$ was
derived in [7] for different (isotropic) dynamical rules.

An expansion for fixed $b>0$ yields $$E(u) = {bp \over
1-u}+{1-b^2 \over \left(1-a^2\right)b^3p}
-{2\left(1-b^2\right)\left(2-b^2p\right) \over
\left(1-a^2\right)^{3/2}b^5p^2} \sqrt{1-u\,}+{\cal O}(1-u)
\;\; . \eqno(21)$$ The leading term corresponds to the
constant contribution $c(t) = bp + \ldots$ which is
expected for the majority species. In fact, expansions near
$u=1$ are nonuniform in the limits $b\to 0^+$ and $b\to
0^-$; here we used for the first time the fact that the
majority species is $A$. The approach to the constant
asymptotic density is given by the third term, $$c(t) -bp
\approx {\left(1-b^2\right) \left(2-b^2p\right) \over
\sqrt{\pi} \, b^5 p^2 \left(1-a^2\right)^{3/2} t^{3/2}}
\;\; . \eqno(22)$$ This difference is just the density of
the minority species $B$. As before, the anisotropy
dependence of this leading-order power-law correction is
fully absorbed in the diffusion rate, while the exponent is
consistent with the results of [7].

It is of interest to explore the nonuniform behavior near
$b=0$ within the crossover scaling formulation. The
appropriate scaling combination turns out to be
proportional to $b/(1-u)^{1/4}$, as determined by
inspection of various limiting expressions. It proves
convenient to absorb certain constants into the precise
definition of the scaling combination $\sigma$, $$ \sigma =
\sqrt{p}\,\left(1-a^2\right)^{1/4}b/(1-u)^{1/4} \;\; .
\eqno(23) $$ In the double-limit $b \to 0$ and $u \to 1^-$,
taken with fixed values of $\sigma$, we obtain the scaling
relation $$ E(u) \approx p^{-1} \left(1-a^2\right)^{-1}
b^{-3} R(\sigma) \;\; , \eqno(24) $$ where the scaling
function $R$, analytic at $\sigma = 0$, can be derived
exactly, $$ R(\sigma) = {\sigma^3
\left(\sigma+\sqrt{4+\sigma^2}\right)^2 \over 4
\sqrt{4+\sigma^2}} \;\; . \eqno(25)$$

For $\sigma \ll 1$ the following expansion applies, $
R(\sigma)={1 \over 2}\sigma^3+{1 \over 2}\sigma^4+{\cal
O}\left(\sigma^5\right)$.  The leading term here reproduces
the first term in (19). The latter was the limiting form
for $u \to 1$ at $b=0$.  The second term in (19), however,
is not of the form $\sim b^{-3}\sigma^4$. Corrections to
the leading scaling behavior correspond to this term in the
$b=0$ expansion (19).

In the opposite limit, $\sigma \to +\infty$, we get the
expansion $ R(\sigma)=\sigma^4+1-4 \sigma^{-2}+{\cal
O}\left(\sigma^{-4}\right)$.  The leading term here
reproduces the first term in (21); the limit $\sigma \to
+\infty$ corresponds to $u \to 1$ at fixed small positive
$b$.  Interestingly, the next two terms in (21) are also
reproduced in their small-$b$ form by the next two terms
here. The second term yields
$1/\left[\left(1-a^2\right)b^3p\right]$ in $E(u)$.  The
third term in (21) is reproduced with the numerator 4 which
is the small-$b$ limiting value.

The scaling description provides a uniform limiting
approximation in the double-limit $b\to 0$ and $u\to 1$.
Specifically, the region of nonuniform behavior near $b=0$
is exploded by the large factor $\sim (1-u)^{-1/4}$. In
terms of $\sigma$, the behavior is smooth and well defined.
For instance, the result (25) applies equally well for
$\sigma<0$ which corresponds to $A$ becoming the minority
species. The limit of $u\to 1^-$ at small fixed $b<0$ is
described by the limit $\sigma \to -\infty$. The
appropriate expansion takes the form $ R(\sigma)=-1+4
\sigma^{-2}+{\cal O}\left(\sigma^{-4}\right)$, similar in
structure to the $\sigma \to +\infty$ expansion but without
the constant-density first term.

In summary, our exact results for the leading-order
large-time particle densities of reaction-diffusion models
in $1D$ show expected universal power-law behaviors.
Anisotropy of hopping has no effect on the universality
class of the models studied; it can be absorbed in the
diffusion constant.  Finite-time results are more sensitive
[6] to the value of the anisotropy parameter $a$; they are
cumbersome to derive and of less interest than the
leading-order expressions. One interesting exception is the
relation (14) which holds in our synchronous-dynamics
models.

\vfil\eject

\noindent {\bf REFERENCES}

\

{\frenchspacing

\item{[1]} T. Liggett, {\sl Interacting Particle Systems\/}
(Springer-Verlag, New York, 1985); V. Kuzovkov and E.
Kotomin, Rep. Prog. Phys.  {\bf 51}, 1479 (1988); V.
Privman, in {\sl Trends in Statistical Physics}, in print
(Council for Scientific Information, Trivandrum, India).

\item{[2]} D. Toussaint and F. Wilczek, J. Chem. Phys.
{\bf 78}, 2642 (1983); K. Kang and S. Redner, Phys. Rev.
Lett. {\bf 52}, 955 (1984); K. Kang, P. Meakin, J.H. Oh and
S. Redner, J.  Phys. A {\bf 17}, L665 (1984); K. Kang and
S. Redner, Phys. Rev. A{\bf 32}, 435 (1985); S. Cornell, M.
Droz and B. Chopard, Phys. Rev.  A{\bf 44}, 4826 (1991); V.
Privman and M.D. Grynberg, J. Phys. A{\bf 25}, 6575 (1992);
B.P. Lee, J. Phys. A{\bf 27}, 2533 (1994).

\item{[3]} M. Bramson and D. Griffeath, Ann. Prob. {\bf 8},
183 (1980); D.C. Torney and H.M. McConnell, J. Phys. Chem.
{\bf 87}, 1941 (1983); Z. Racz, Phys. Rev. Lett. {\bf 55},
1707 (1985); A.A. Lushnikov, Phys. Lett. A{\bf 120}, 135
(1987); M. Bramson and J.L. Lebowitz, Phys. Rev. Lett.
{\bf 61}, 2397 (1988); D.J. Balding and N.J.B. Green, Phys.
Rev. A{\bf 40}, 4585 (1989); J.G. Amar and F. Family, Phys.
Rev. A{\bf 41}, 3258 (1990); D. ben-Avraham, M.A. Burschka
and C.R. Doering, J. Stat. Phys. {\bf 60}, 695 (1990); M.
Bramson and J.L. Lebowitz, J. Stat.  Phys. {\bf 62}, 297
(1991); V. Privman, J. Stat. Phys. {\bf 69}, 629 (1992).

\item{[4]} R. Kopelman, C.S. Li and Z.--Y. Shi, J.
Luminescence {\bf 45}, 40 (1990); R. Kroon, H. Fleurent and
R. Sprik, Phys. Rev.  E{\bf 47}, 2462 (1993).

\item{[5]} S.A. Janowsky, Phys. Rev. E, in print.

\item{[6]} V. Privman, J. Stat. Phys. {\bf 72}, 845 (1993);
V. Privman, E. Burgos and M.D. Grynberg, preprint.

\item{[7]} P. Krapivsky, Physica A{\bf 198}, 135 (1993); P.
Krapivsky, Physica A{\bf 198}, 150 (1993).

\item{[8]} V. Privman, A.M.R. Cadilhe and M.L. Glasser,
preprint.

\item{[9]} I.M. Sokolov and A. Blumen, Phys. Rev. E{\bf
50}, 2335 (1994).

\item{[10]} V. Privman, Phys. Rev. E{\bf 50}, 50 (1994).

\item{[11]} H. Takayasu, Phys. Rev. Lett. {\bf 63}, 2563
(1989).

\item{[12]} H. Takayasu, M. Takayasu, A. Provata and G.
Huber, J. Stat. Phys. {\bf 65}, 725 (1991); S.N. Majumdar
and C. Sire, Phys. Rev. Lett.  {\bf 71}, 3729 (1993).

\item{[13]} J.L. Spouge, Phys. Rev. Lett. {\bf 60}, 871
(1988).

}

\bye